\documentclass[11pt]{llncs} 

\usepackage{graphicx} 
\usepackage{listings}
\usepackage{url}
\usepackage{subfigure}
\usepackage{setspace}
\usepackage{wrapfig}
\usepackage{a4wide}

\usepackage{booktabs} 
\usepackage{array} 
\usepackage{paralist} 
\usepackage{verbatim} 
\usepackage{subfig} 

\title{A General Architecture for Heterogeneous Language Engineering and Projectional Editor Support}
\author{Tony Clark}

\institute{School of Science and Technology, Middlesex University, London, UK\\ \email{t.n.clark@mdx.ac.uk}}

\begin{document}

\maketitle

\begin{abstract}
Tool support for language engineering has typically prioritises concrete syntax over abstract syntax by providing meta-languages for expressing concrete syntax and then mapping concrete to abstract structures. Text-based languages are usually specified using a BNF-like language used to generate a syntax-aware editor that includes features such as keyword completion. Similarly, graphical languages are defined using a declarative graphical syntax language, producing an editor that supports features such as shapes, graphs and edges. Projectional editors invert traditional approaches by prioritising abstract over concrete syntax. This paper describes a projectional meta-tool architecture, including general purpose abstract and concrete meta-languages, that uses declarative rules to integrate the syntax and tool support for a range of heterogeneous languages. The architecture has been implemented in Racket and the paper illustrates the architecture with concrete examples.
\end{abstract}

\section{Introduction}

Domain Specific Languages (DSLs) \cite{van2000domain,mernik2005and} are motivated by the need to define languages that match specific use-cases, as opposed to General Purpose Languages (GPLs). Whilst GPLs are usually supported by standard text editors, DSLs, by their nature, often contain a range of more exotic syntax elements that are arguably better supported by syntax-aware editors. 
This has led to the development of a range of technologies to support DSL development and that generate tools for each DSL. Where the DSL is limited to text, languages such as EMFText \cite{heidenreich2009derivation}, MontiCore \cite{krahn2008monticore,krahn2010monticore}, TCS \cite{jouault2006tcs}, and XText \cite{eysholdt2010xtext} and Spoofax \cite{kats2010spoofax}  allow a DSL to be quickly and conveniently defined and the associated tooling generated. These technologies are mainly based on grammar-ware \cite{klint2005toward} that integrate language parsers with editors in order to achieve a workbench. Many of the technologies integrate static and dynamic analysis of the resulting DSL. These technologies have become quite mature and the term Language Workbench \cite{fowler2005language} has been coined to describe this type of engineering tool.

Whilst languages used for programming or scripting tend to be exclusively text-based, modelling languages have included a much wider palette of elements. UML for example, has a number of sub-languages that are based on graphs, but also includes text in the form of OCL and action languages. Relatively few technologies support the definition and tooling of DSLs containing graphical syntax elements. Notable exceptions are Eugenia \cite{kolovos2010taming,kolovos2009raising}, GMF \cite{herrmannsdoerfer2010language}, MetaEdit+ \cite{tolvanen2009metaedit+}. However, these technologies do not support multi-mode languages where text and graphics can be freely mixed.  In additional they do not have intrinsic support for defining semantics, particularly operational semantics, for the languages that are defined, and require the language engineer to step out of the system in order to provide such information.

There has been increasing attention to mixing graphical and textual notations \cite{andres2008domain,schneider2011integrating,engelen2010integrating,scheidgen2008textual}. A recent model-based approach to mixing text and graphical languages is described in \cite{atkinson2013harmonizing} that uses projectional editing techniques over a model. Whilst most of the reported work agrees on the general principles and propose approaches, this paper appears to be the first to propose concrete meta-language that integrates with a general architecture for projectional editing in order to integrate textual and graphical languages.

Language engineering technologies such as those above tend to focus on concrete-syntax as the primary consideration. The language is often defined as a grammar or a collection of graphical elements that are then mapped to abstract-syntax structures. This is usually a one-to-one relationship. Projectional editors invert the traditional syntax dominance to promote abstract-syntax over concrete-syntax. The benefit of this is that the language definition does not dictate a single user experience since a single abstract-syntax can be mapped to multiple concrete-syntaxes depending upon mode, or execution state, or viewpoint.

The disadvantage of the projectional approach is that the supporting technology is less mature and less widespread. A small number of implementation platforms exist, such as MPS \cite{voelter2012language,voelter2010language} and Intentional Software \cite{simonyi2006intentional}. Whilst both of these technologies have been several years in development, there is no basic definition of a {\it projectional editor} or a framework for exploring variations.

This paper presents a simple architecture for a projectional editor and identifies several key features that should be supported including syntax representation, mappings and identity management. It goes further by proposing a simple meta-language that can be used to populate and drive such an architecture. The meta-language uses pattern-directed rules to transform syntax trees. It is based on the transformation rules of Stratego/XT \cite{bravenboer2008stratego}, and is a distant cousin to model and language transformation engines such as ATL \cite{jouault2006atl} and TXL \cite{cordy2002source} but extends these in terms of expressive power, and proposes two different categories of mapping: {\it transformations} and {\it reductions}. Both the architecture and meta-language have been implemented in Racket and the approach is demonstrated through a number of concrete running examples.

\section{Example}

\label{sec:game}

\begin{figure}[t]
\centering
\includegraphics[width=0.6\textwidth]{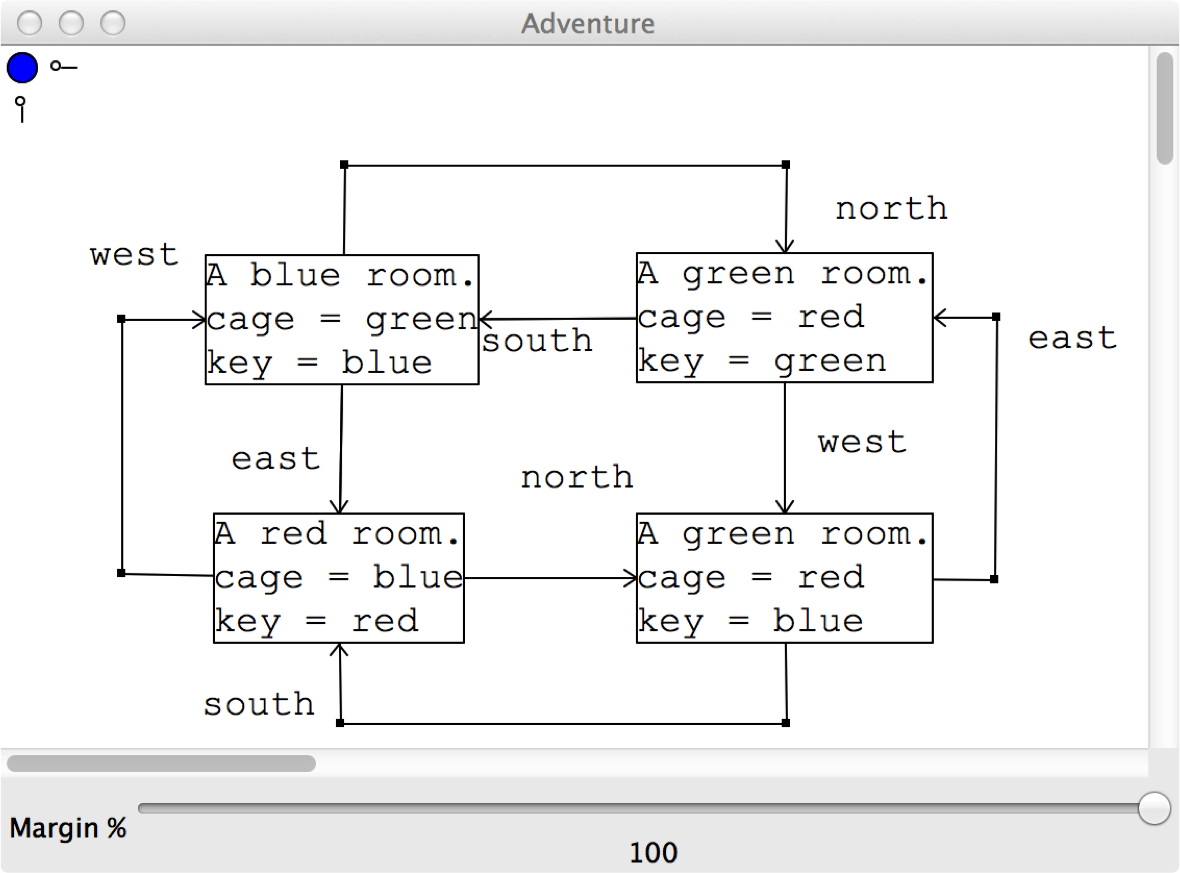}
\caption{Defining the Game}
\label{fig:dungeon}
\end{figure}

\begin{figure}[t]
\centering
\subfigure[Starting State]{\includegraphics[width=0.4\textwidth]{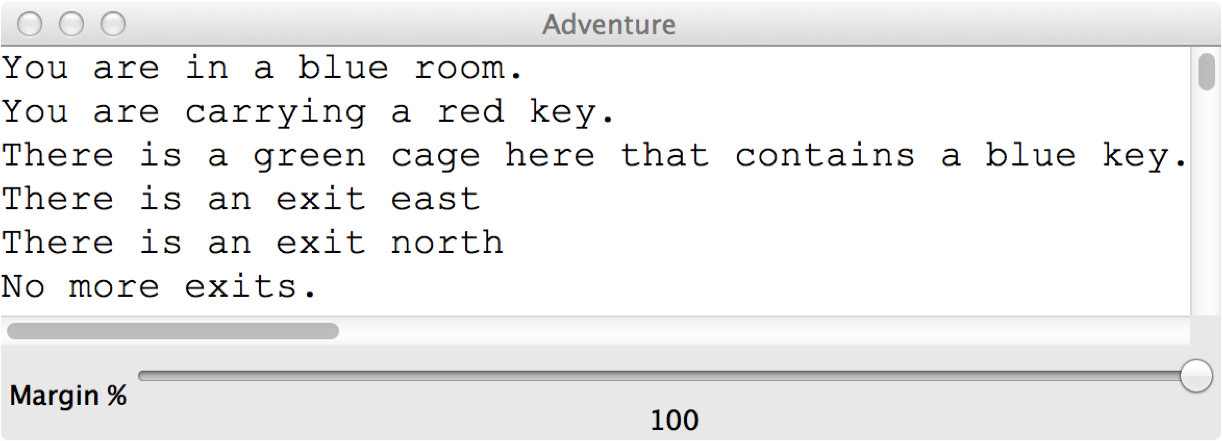}}
\subfigure[n]{\includegraphics[width=0.4\textwidth]{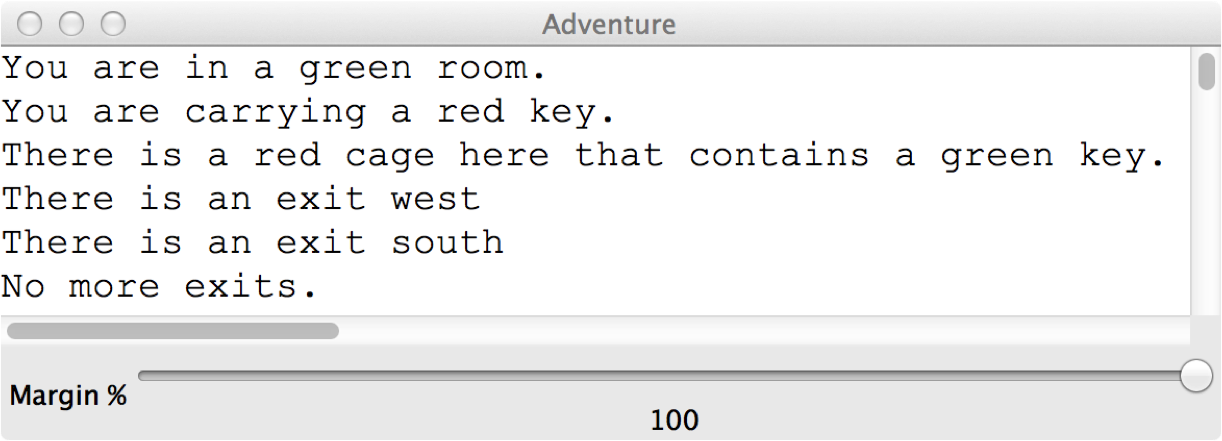}}
\subfigure[u]{\includegraphics[width=0.4\textwidth]{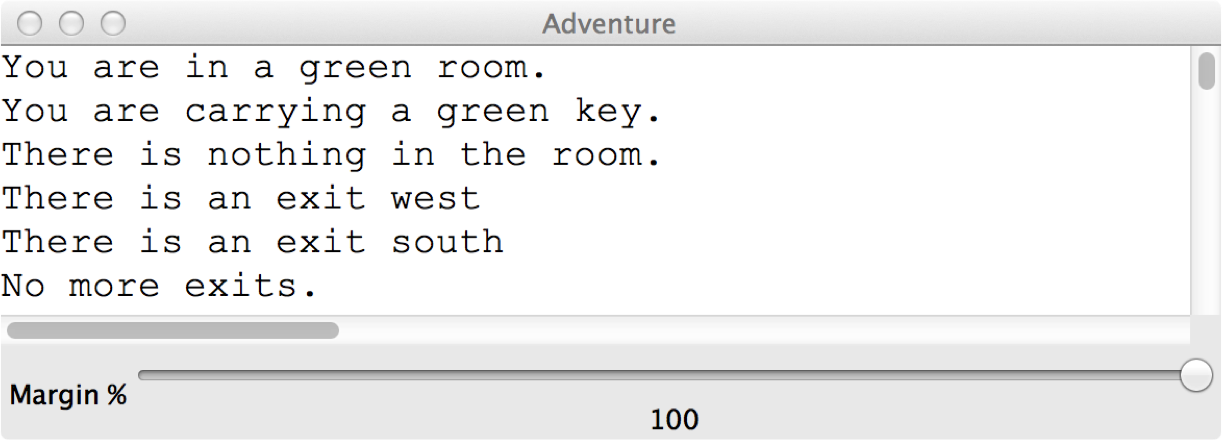}}
\subfigure[s]{\includegraphics[width=0.4\textwidth]{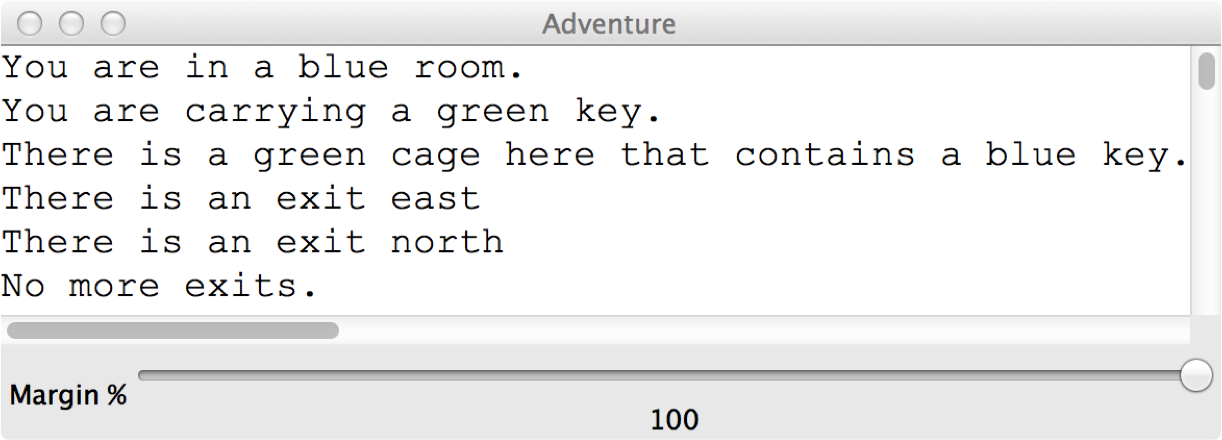}}
\caption{Playing the Game}
\label{fig:game}
\end{figure}

Consider a game that involves a collection of rooms that are connected by corridors. A room is either empty or contains a locked cage. The cage is painted red, green or blue. Inside the cage is a painted key. A key can be used to unlock a cage of the same colour and get the key inside. The player starts off in a room with a red key. The aim of the game is to visit all the rooms and unlock all the cages.
Figure \ref{fig:dungeon} shows the definition of a dungeon using the language editor for game construction. Rooms are created as nodes and corridors as labelled edges. The text in a room-node shows the colour of the room, the colour of a cage and the colour of the key in the cage. The blue dot at the top-left corner of the tool is used to access a room-creation menu. Edges between room-nodes are created by dragging the mouse from a source node to the target (a menu is used to select a direction label). When a room-node is created, its colour and contents are unset. The mouse is used to select from pre-defined colours for the room, cage and key. 

The language operates in two modes: creation and play, it is possible to switch between the modes by pressing {\tt p} and {\tt c} on the keyboard. Figure \ref{fig:game} shows play mode. The player starts in the blue room with a red key. The player makes a move by pressing the first letter of the direction on the keyboard. Since th eplayer does not have a green key they must move from the starting room; they press {\tt n} to go north and arrive at a green room with a red cage. The player can open the cage since their key matches the cage colour. This is done by pressing {\tt u} on the keyboard. Finally, the player goes back south.

\begin{figure}[t]
\centering
\subfigure[Room Creation]{\includegraphics[width=0.3\textwidth]{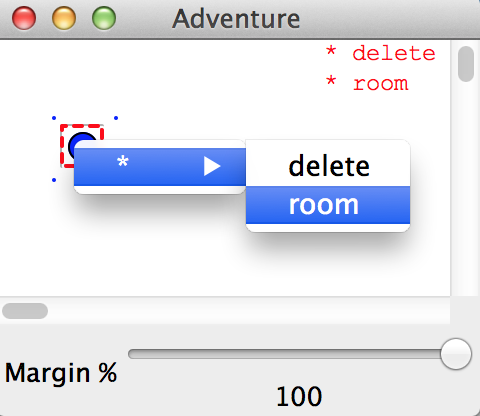}}
\subfigure[Setting Room Colour]{\includegraphics[width=0.3\textwidth]{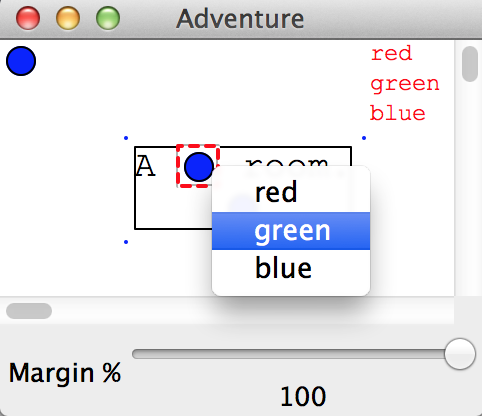}}
\subfigure[Edge Creation]{\includegraphics[width=0.3\textwidth]{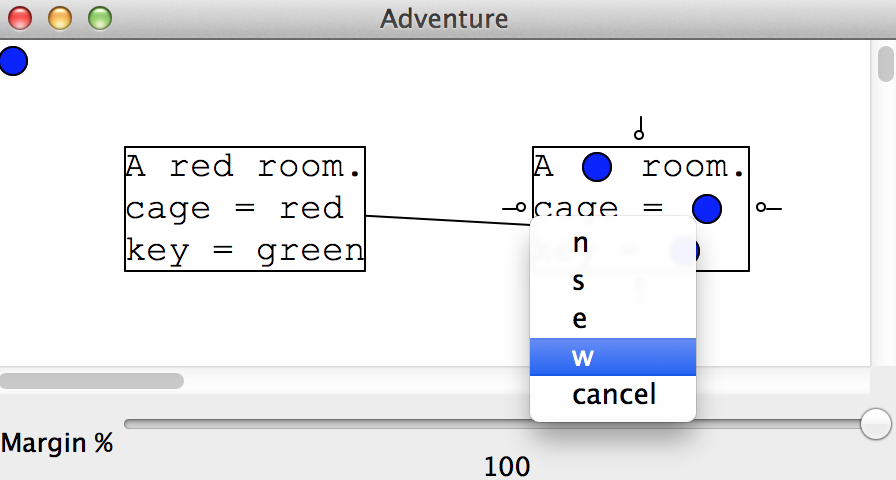}}
\caption{Editor Interactions}
\label{fig:editor_interaction}
\end{figure}

The game shows a number of features of the projectional editor. Interaction with the language can be moded; in this example there are two modes, but in general there can be any number. Figure \ref{fig:dungeon} shows that the abstract syntax can be projected on to graphs and text. In addition, language features can be created by menus made available as blue-dots. Figure \ref{fig:dungeon} shows a blue dot that is used to create room-nodes, but in general a language may offer many different types of item. Figure \ref{fig:game} shows that the state of the game is projected to become formatted text. Figure \ref{fig:editor_interaction} shows how the editor that is generated from the language definition supports creation of language elements: (a) creation of a new room element; (b) selection of a room colour; (c) selection of a type of edge between rooms.

\section{An Architecture for Projectional Editors}

\label{sec:architecture}

\begin{figure}[t]
\centering
\includegraphics[width=0.75\textwidth]{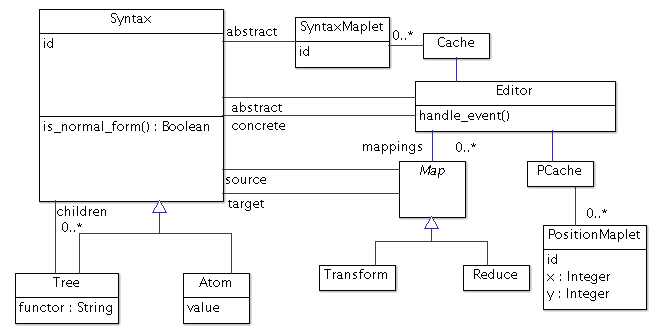}
\caption{The Editor Architecture}
\label{fig:editor_architecture}
\end{figure}

A projectional editor provides facilities for defining and constructing abstract syntax, reducing the abstract syntax to concrete syntax and then mapping user actions that are applied to the concrete syntax into transformations applied to the abstract syntax. The characteristic feature of the closure of the loop:
\begin{lstlisting}[deletekeywords={reduce,abstract,transform},numbers=none]
loop {
  concrete := reduce(abstract);
  display(concrete);
  wait(concrete-event) {
    abstract := transform(concrete-event,abstract);
  }
}
\end{lstlisting}
Figure \ref{fig:editor_architecture} shows the proposed architecture for an editor that supports such a loop. Both concrete and abstract syntax can be implemented as a simple tree structure where the type of the structure is encoded as a functor at the root of each sub-tree. 
Abstract syntax will depend on the particular language being defined, however the concrete syntax must conform to a particular interface that is supported by the editor in order that it can display the syntax and can listen for user events. Therefore, {\tt Syntax} defines a predicate {\tt is\_normal\_form()} that is true when the syntax is in a format suitable for the user.

The loop shown above must map concrete events to actions that can be applied to the abstract syntax. In addition, each time round the loop, the abstract syntax produces a {\it different} concrete syntax structure. In both cases it must be possible to keep track of the association between abstract and concrete syntax elements. In the first case the association is used to apply the action to the correct abstract syntax element, and in the second case the user may have changed the concrete syntax, for example by moving graph nodes, and the new concrete syntax element must be associated with a current position.

The architecture shown in figure \ref{fig:editor_architecture} uses identities on syntax elements in order to implement a mapping between abstract and concrete syntax. An fresh identity is coined when a new abstract element is created and may be passed on to individual concrete elements via the mapping. The editor maintains two caches. The first associates identities with abstract elements in order to support the event to action mapping. The second associates concrete identities with positional information.

The editor provides two types of mapping. The first is a transformational mapping that is used to change the abstract syntax tree. An example of such a transformation is the creation of a new node or editing a string. There is no requirement that the result of such a mapping is in normal form. The second type of mapping is a reduction that takes abstract syntax and produces syntax {\tt t} such that {\tt t.is\_normal\_form()}.

The architecture described above has been implemented as a tool in Racket. The editor is driven by a meta-language that defines the structure of the syntax trees and implements the mappings using pattern matching over trees. The rest of this paper describes the meta-language and provides a number of examples of its use that exemplify key features. The paper concludes with the implementation of the game described in section \ref{sec:game} and a discussion of possible extensions to the architecture.

\section{A Meta-Language Definition Language}

\begin{figure}[t]
\centering
\begin{minipage}[t]{\textwidth}
\begin{minipage}[t]{0.5\textwidth}
\begin{lstlisting}[deletekeywords={string,char,exp},morekeywords={*,or,text,str,dots,case},numbers=none]
lang = (deflang id
         (abstract (id g-element ...))
         (locals local-def ...)
         (transform rule ...)
         (reduce rule ...))
g-element = id | str 
          | (id g-element ...)
          | (* g-element ...)
          | (or g-element ...)
local-def = (id exp)
          | ((id pattern ...) exp)
rule      = (pattern exp)
pattern   = id | string | int | char | bool
          | (id pattern ...)
          | ((id pattern ...) pattern ...)
          | pattern dots
exp       = id | string | int | char | bool
          | (id exp ...)
          | ((id exp ...) exp ...)
          | exp dots
          | (case exp rule ...)
\end{lstlisting}
\end{minipage}
\begin{minipage}[t]{0.4\textwidth}
\begin{lstlisting}[deletekeywords={string,char},morekeywords={menu,left,right,top,bottom,centre,align,chars,*,or,border,ellipse,indent,box,tree,underline,vbox,hbox,rectangle,image,graph,types,node,edge,label,arrow,none,source,target,thumbnail,dots,tab,case,nl,seq},numbers=none]
nf = (nl) | string | int | bool | char
   | (font +/- nf)
   | (seq nf ...)
   | (tab nf)
   | (indent int nf)
   | (box bb bm (pos nf) ...)
   | (vbox bb bm (pos nf) ...)
   | (hbox bb bm (pos nf) ...)
   | (ellipse int int int int bool bool)
   | (rectangle int int int int bool bool)
   | (image int int string)
   | (underline nf)
   | (chars string)
   | (thumbnail int int int int)
   | (tree nf ...+)
   | (graph 
       (edge-types (id (id) (id)) ...) 
       g-node ... 
       g-edge ...)
+/-    = (+) | (-) 
bb     = (border int)
bm     = (menu (id (id nf ...)) ...)
pos    = left | right | top | bot | centre | align
g-node = (node (id) nf nf)
g-edge = (edge nf nf e-dec nf e-dec (label end nf))
e-dec  = (arrow) | (none)
end    = (source) | (target)
\end{lstlisting}
\end{minipage}
\end{minipage}
\caption{The Language-Definition Language}
\label{fig:meta_language}
\end{figure}

The language architecture described in section \ref{sec:architecture} is defined with respect to a language definition that describes the structure of the abstract syntax tree, the transformation rules and the reduction rules. In addition a language definition may contain a collection of local value and function definitions. A language definition is written in a meta-language that is introduced by this section. The language definition is described in overview and the key features of the language are given as a series of examples that follow.

Figure \ref{fig:meta_language} defines the meta-language. A language defined using {\tt deflang} introduces the abstract syntax in the {\tt abstract} clause, the transformation rules in the {\tt transform} clause, and the reduction rules in the {\tt reduce} clause. The local identifiers introduce by {\tt locals} are scoped over the rules. An abstract-clause is a tree-structure definition, including alternatives. It is defined using mutually recursive named abstract-rules where {\tt *} is used to express 0 or more repetitions and {\tt or} is used to express alternatives. An editor starts with respect to a named abstract-rule causing the abstract-syntax tree to be created as an instance of the clause, up to the point where alternatives or repetition is encountered. At that point the editor must wait for user input in the form of menu selection in order to choose between the alternatives, to add another instance of a repeated clause or to delete a repeated clause. For example, given the abstract definition: 
\begin{lstlisting}[numbers=none]
(abstract 
  [tree (node (* (or leaf tree)))]
  [leaf (data str)])
\end{lstlisting}
and a starting clause {\tt tree}, the editor will generate abstract syntax for {\tt (node h)} where {\tt h} is a {\it hole} requiring user input to produce another child. If the user opts to produce another child then the result is another hole that requires the user to select between a leaf or a tree. Selecting a leaf will replace the hole with a child that has the form {\tt (data s)} for an editable string {\tt s}. Selecting a tree, will cause the hole to be replaced with a new {\tt (node h)} structure, causing the process to be replicated at the child level.

The abstract-clause provides guidance to the editor on the user-assisted development of the abstract syntax structure. However, it is not a type definition, since the transformation rules are free to replace any element of the abstract syntax with structures that are not defined within the cause.

Both the transformation and reduction rules are defined in terms of {\it rules}. A rule has a pattern and an expression. When applied to an abstract syntax tree, the rules are tried in turn: if their pattern matches the current structure then the expression is evaluated to produce a replacement structure. In both cases, the rule-sets are applied using a top-down, left-to right strategy. If the rules match then the matching sub-tree is replaced with the freshly constructed tree and the process is repeated from the root. The process terminates when there is no sub-tree that matches any rule-pattern.

If a pattern matches a syntax-tree then the result is a collection of identifier bindings that are used in the corresponding expression evaluation. If an identifier is encountered twice in the same pattern then it must be bound to the equal values, where values with the same structure are equal even if they have different identities. A pattern {\tt (f x y z)} will match any tree with the same functor {\tt f} and where the corresponding sub-trees match. A pattern {\tt ((f i) x y z)} will match as before, but will also match {\tt i} against the identity of the tree. When an expression {\tt (f x y z)} is performed, a tree with a new identity is created. The expression {\tt ((f i) x y z)} creates a tree with the designated identity {\tt i} thereby allowing tree identities to be propagated by rules.

Syntax trees may have multiple children. The pattern {\tt (f p ...)} matches any tree that has a root-functor {\tt f} and for which each child matches {\tt p}. Each variable in {\tt p} may be matched multiple times in which case they may be used multiple times in expressions by adding {\tt ...} after an expression that refers to them. For example, the following rule: {\tt\small
[(node c1 ... (data "x") c2 ...) (node c1 ... c2 ...)]}
may be used to remove a child from a node. Local definitions might be used to remove all children of a particular form:
\begin{lstlisting}[numbers=none]
(locals
  [(remove-xs n)
   (case n
     [(node c1 ... (data "x") c2 ...)     (remove-xs (node c1 ... c2 ...))]
     [(node c ...)                        (node (remove-xs (node c)) ...)]
     [(data s)                            (data s)])])
(transform
  [(node c ...) (remove-xs (node c ...))])
\end{lstlisting}
The reduction rules of a language are used to produce a normal-form that represents the concrete syntax to be displayed to the user. Figure \ref{fig:meta_language} shows the definition {\tt nf} of normal-forms. In most cases these are trees with designated functors, for example {\tt (box (border 1) (centre "x"))}. In some cases it is important that the identity of the normal-form syntax-tree is consistent for each reduction. Where this is necessary, the identity {\tt i} can be set as above: {\tt ((box i) (border 1) (centre "x"))}.

Normal-forms describe how they should be displayed on the screen. In addition, boxes contain menus that appear when the box is selected. Each menu element contains a message that is sent to the current abstract syntax tree when selected. The following list provides an overview of normal-forms: {\tt nl} produces a new-line; {\tt seq} is used to display a sequence of elements; {\tt tab} and {\tt indent} are used to control text formatting; {\tt ellipse}, {\tt rectangle}, {\tt image} and {\tt thumbnail} are used to display shapes; {\tt chars} creates a text editor. A normal-form created using {\tt tree} contains a root normal-form and any number of children.

Boxes contain elements and are created using {\tt box} in which case the elements are positioned relative to the top-left corner of the box, {\tt hbox} where the elements are listed horizontally, or {\tt vbox} where the elements are listed vertically. Boxes may have outlines and have menus as discussed above. Boxes are selectable elements and therefore it is important that they have a consistent identity for each reduction.

Graphs are created using {\tt graph} and contain a collection of edge types, nodes and edges. The edge types determine whether edges can be drawn between designated node types when the user drags the mouse from a source node to a target node. Nodes are created as {\tt ((node i) (t) j nf)} where {\tt i} should be consistent over reductions and allows the editor to cache the node position, {\tt t} is the node type, {\tt j} is the identity of the abstract syntax tree that has been mapped to the node, and {\tt nf} is the concrete-syntax to display on the node. Edges should also have consistent identities and include information about end decorations and labels.

The following sections provide examples of the use of the meta-language. In each case the example has been implemented in the editor, although the definitions have been reduces slightly by omitting certain details where this should not cause confusion.

\section{Language Definition and Pattern-Based Reduction}

\begin{figure}[t]
\centering
\begin{tabular}[b]{lcr}
\begin{minipage}[b]{0.4\textwidth}
\begin{lstlisting}
(deflang DNA
  (abstract
   [DNA (gene (* letter))]
   [letter (or (a) (c) (t) (g))])
  (reduce
   [(gene letter ...) (seq letter ...)]
   [(a) "A"]
   [(c) "C"]
   [(t) "T"]
   [(g) "G"]))
\end{lstlisting}
\end{minipage}
&
\hspace{1cm}
&
\includegraphics[width=0.3\textwidth]{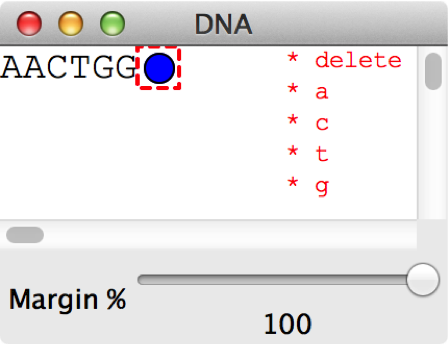}
\end{tabular}
\caption{Sequences of DNA}
\label{fig:dna}
\end{figure}

Figure \ref{fig:dna} shows a simple example of a language definition and the editor that is produced. The abstract syntax structure that is shown on the right is {\tt (gene (a) (a) (c) (t) (g) (g))}. Since the abstract-syntax defines a gene tree to contain {\tt * letter} the blue circle denotes a hole that can be selected using the mouse to add more letters. The red menu on the right shows the keyboard short-cuts that can be selected by pressing the key corresponding to the first letter of the option.

The concrete syntax produced by the {\tt reduce} rules is a sequence of strings: {\tt (seq "a" "a" "c" "t" "g" "g" h)} where {\tt h} is a hole. The editor displays sequences of elements horizontally on the screen. The hole is used to capture mouse events that allow the user to create new letters.

\section{Boxes and Trees}

\begin{figure}[t]
\begin{center}
\begin{lstlisting}
(deflang boxes
  (abstract [root (root (boxes) tree)] [tree (or (tree str (* tree)) (leaf str))])
  (transform
   [(send (root _ t) (key-pressed _ #\t)) (root (tree) t)]
   [(send (root _ t) (key-pressed _ #\b)) (root (boxes) t)])
  (reduce
   [(root (boxes) t) (tree->boxes t)]
   [(root (tree) t)  (tree->tree t)]
   [(tree->boxes (tree data child ...))
    (hbox (outline 1) (centre data) (align (vbox (outline 1) (align (tree->boxes child)) ...)))]
   [(tree->boxes (leaf d)) d]
   [(tree->tree  (tree data child ...)) (tree data (tree->tree child) ...)]
   [(tree->tree  (leaf d)) d]))
\end{lstlisting}
\end{center}
\caption{Trees Displayed in Two Modes}
\label{fig:trees}
\end{figure}

Concrete syntax needs to structure information so that it can be presented to users in a suitable way. The previous section showed how {\tt seq} is used to linearise information. Boxes can be used to define groups of concrete syntax elements that to organise them horizontally and vertically. Boxes conveniently nest so that large scale structures can be organised. Another way to organise structured information is to use trees. A tree has a data element at the root and contains a sequence of elements as children. Trees can be nested to any depth. 

Decision trees can be organised using nested boxes or trees. Consider the following abstract syntax tree:
\begin{lstlisting}[numbers=none]
(root (tree)
  (tree "hair?"
    (tree "legs < 5?" (leaf "mammal") (leaf "insect"))
    (tree "feathers?" 
      (leaf "bird") 
      (tree "tail?"
        (tree "legs < 2?"
          (leaf "fish")
          (tree "legs < 6?" (leaf "reptile") (leaf "shellfish")))
        (tree "legs < 5?" (leaf "frog") (leaf "insect"))))))
\end{lstlisting}
The tree is used to determine the type of an animal. The idea is that the decision making process starts at the root with the question {\it does the animal have hair?}. If the answer is yes, then move on to the first sub-tree, otherwise move to the second sub-tree. If the process ever reaches a leaf, the animal has been categorised.

The tree can be represented using nested boxes by representing a tree-node as a horizontal box with the question on the left and the sub-trees on the right. Each sub-tree is listed one above the other so that if the answer to the question is {\it yes} then the first sub-tree is selected.

Figure \ref{fig:trees} shows the definition of the decision tree language. The root node of the abstract syntax contains an element {\tt (boxes)} or {\tt (tree)} that is used to mode the reduction rules. The reduction rule {\tt tree->boxes} is used to map the abstract syntax to nested boxes whereas the rule {\tt tree->tree} is used to produce nested trees.

Boxes may specify an outline using the optional {\tt (outline n)} element where {\tt n} is the width of the line. Each element of a box must be supplied with a position. In the example the question is centred the {\tt hbox} whereas the sub-trees are aligned.

The definition of the reduction rules in figure \ref{fig:trees} provides an example of how rules process recursively defined structures. Notice that the definition of rules {\tt tree->boxes} and {\tt tree->tree} uses both of these functors in the expression part of the rule. This should not be interpreted as a recursive {\it call} to the rule: the expression is creating a new abstract syntax structure with the supplied functor. The operational semantics of rule matching means that if a rule is applied to a tree then all rules are re-applied to the root of the tree. Therefore the effect is a recursive call to the rule.

Figure \ref{fig:trees}  shows how transformation rules are used to process events. When the user presses a key {|tt k}, the message {\tt (send s (key-pressed i k))} is sent to the abstract syntax tree. In effect this is done by updating the tree to become a new tree and then running the mappings. Transformation rules can make changes to the abstract syntax tree managed by the editor and therefore is used to consume messages sent to the tree. The transformation rules in the example, show that the keys are consumed and cause the mode on the root of the abstract syntax tree to be changed.

\begin{figure}[t]
\centering
\subfigure[Displaying Decisions as a Tree]{\includegraphics[width=0.45\textwidth]{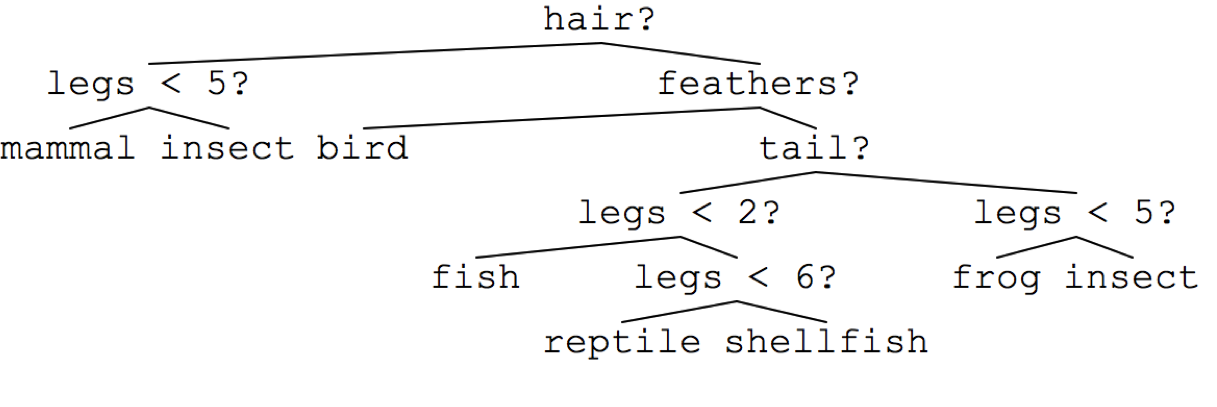}}\qquad
\subfigure[Displaying Decisions as Nested Boxes]{\includegraphics[width=0.45\textwidth]{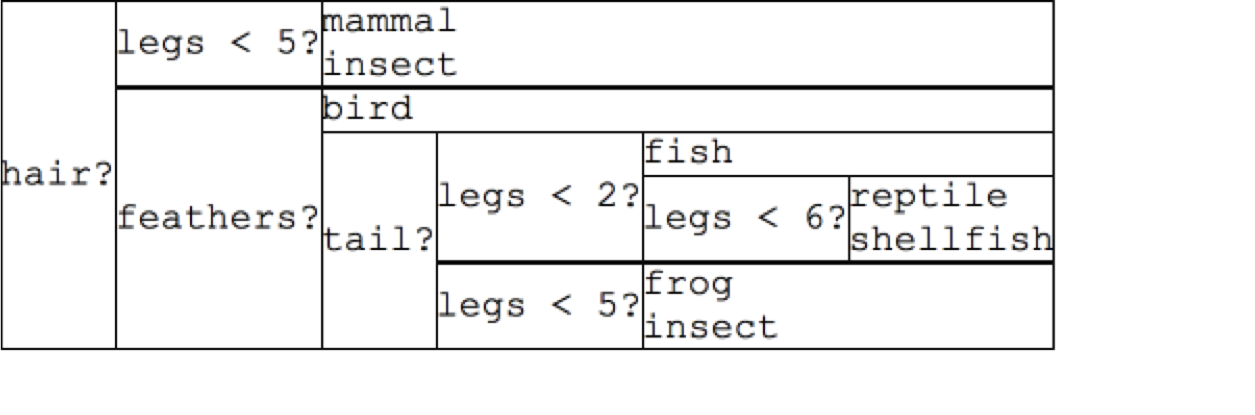}}
\caption{Displaying a Decision tree}
\label{fig:displaying_a_decision_tree}
\end{figure}

Figure \ref{fig:displaying_a_decision_tree} shows two screenshots of the editor. The first shows the example tree where the mode is {\tt (tree)} and the second where the mode is {\tt (boxes)}. The two modes are toggled using the {\tt t} and {\tt b} keys.

\section{Transformations and Evaluation}

\begin{figure}[t]
\begin{center}
\begin{lstlisting}[deletekeywords={apply,str,exp,subst,first,second,eval}]
(deflang lambda-calculus
  (abstract
   [exp (or 
         (const str) 
         (pair exp exp) 
         (ident str) 
         (apply exp exp) 
         (lambda str exp))]) 

  (transform
   [(send t (key-pressed _ #\e))         (eval-step t)]

   [(eval-step (lambda arg exp))         (lambda arg (eval-step exp))]
   [(eval-step (pair exp1 exp2))         (pair (eval-step exp1) (eval-step exp2))]
   [(eval-step exp)                      (eval exp)]

   [(eval (apply (lambda arg body) exp)) (subst exp arg body)]
   [(eval (apply exp1 exp2))             (apply (eval exp1) exp2)]
   [(eval exp)                           exp]

   [(subst new old (const k))            (const k)]
   [(subst new old (pair first second))  (pair (subst new old first) (subst new old second))]
   [(subst new old (ident old))          new]
   [(subst _   old (ident name))         (ident name)]
   [(subst new old (apply exp1 exp2))    (apply (subst new old exp1) (subst new old exp2))]
   [(subst new old (lambda old exp))     (lambda old exp)]
   [(subst new old (lambda arg exp))     (lambda arg (subst new old exp))])

  (reduce
   [((hole _) h)  h]
   [(ident s)     s]
   [(const k)     k]
   [(pair e1 e2)  (tree "." e1 e2)]
   [(apply e1 e2) (tree "@" e1 e2)]
   [(lambda i e)  (tree (hbox (fixed) (align "<*$\lambda$*>") (align i)) e)]))
\end{lstlisting}
\end{center}
\caption{$\lambda$-calculus Evaluation}
\label{fig:lambda-definition}
\end{figure}

Transformation rules make changes to the abstract syntax tree before it is reduced. This feature of the approach can be used to implement an operational semantics for a language. Consider a simple $\lambda$-calculus and a step-wise reduction strategy. A convenient concrete representation for such a calculus is a tree that is rooted either at an application node, at a $\lambda$, or at an atom (identifier or constant). In order to make the language more useful we also add pairs. 

A $\lambda$-expression is in a {\it normal-form} if it contains no applications. An evaluation step performs a single application producing a new expression. The calculus is defined as a language in figure \ref{fig:lambda-definition}.
\begin{figure}[t]
\begin{minipage}{0.5\textwidth}
\begin{lstlisting}[numbers=none]
(lambda "f" 
  (apply 
    (lambda "x" 
      (apply 
        (ident "f") 
        (apply (ident "x") (ident "x")))) 
    (lambda "x" 
      (apply 
        (ident "f") 
        (apply (ident "x") (ident "x"))))))
\end{lstlisting}
\end{minipage}
\begin{minipage}{0.5\textwidth}
\includegraphics[width=\textwidth]{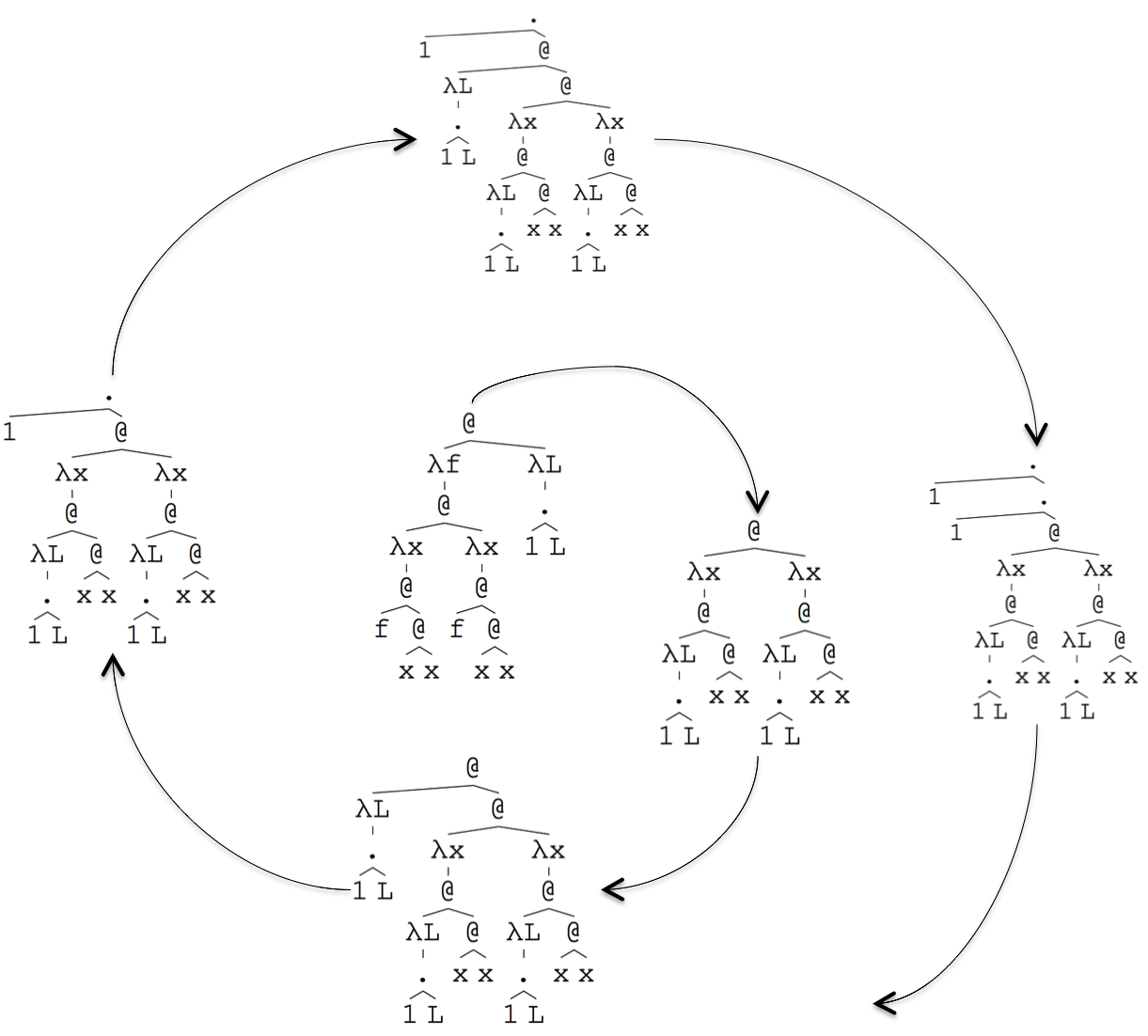}
\end{minipage}
\caption{Definition of {\tt Y} and evaluation of {\tt Y($\lambda$ L. .(1,L))}}
\label{fig:ones}
\end{figure}
A $\lambda$-calculus provides support for recursive definitions using the {\tt Y}-combinator. This is a function that can be applied to a function and return the fixed-point of the function, {\it i.e.}, {\tt f(Y(f))=Y(f)}. Suppose that we define a function {\tt f = $\lambda$ L. (1,L)}, then {\tt Y(f)} is the infinite sequence of {\tt 1}s: {\tt (1,(1,(1,...)))}.
Figure \ref{fig:ones} shows the definition of {\tt Y} as abstract syntax on the left and editor screen-shots starting with {\tt (apply Y (lambda "L" (pair (const "1") (ident "l"))))}. The starting state is shown in the centre of the spiral and each subsequent state is an evaluation-step. The execution shows how the {\tt Y}-combinator gradually emits {\tt 1}s on the left-hand side of the tree.

\section{Graphs}

Many different languages lend themselves to concrete representations in the form of graphs. Class diagrams place classes on nodes and associations on edges. State machines place states on nodes and transitions on edges. Maps show locations on nodes and routes between locations on edges.

The meta-language provides support for graphs by translating abstract syntax to a normal-form:
{\tt (graph (edge-types edge-type ...) node ... edge ...)}
Edge-types are defined to control what happens when a user drags an edge between nodes. Graph nodes are created with a designated type and not all types of node can be connected with edges; furthermore, the same two node types may be connected with different types of edge. For example, on a use-case diagram the edge-type {\it uses} holds between a node of type {\it actor} and a node of type {\it use-case} but not vice versa. In addition, edges of types {\it extends} and {\it includes} may both hold between the same nodes of type {\it use-case}.

Both edge-types and node-types can be any value, but are conveniently expressed as 0-ary trees whose functor designates the type. Suppose that we want to create a use-case graph with a single actor and two use-cases. The actor uses both use-cases and one use-case includes and extends the other (for the sake of the example). The graph can be expressed as follows (assuming appropriate identities and displays):
\begin{lstlisting}[numbers=none]
(graph
  (edge-types
    (uses     (actor)    (use-case))
    (includes (use-case) (use-case))
    (extends  (use-case) (use-case)))
  (node (actor)    id1 actor-display)
  (node (use-case) id2 use-case-display)
  (node (use-case) id3 use-case-display)
  (edge (uses)     id1 (none) id2 (arrow))
  (edge (uses)     id1 (none) id3 (arrow))
  (edge (includes) id2 (none) id3 (arrow) (label (target) "<<includes>>"))
  (edge (includes) id2 (none) id3 (arrow) (label (target) "<<extends>>")))
\end{lstlisting}
Graphs can contain arbitrarily complex displays on the nodes and the edge-labels. As an example of this consider the case where an abstract syntax structure is defined to be recursive in terms of the data on the node, {\it i.e.}, graphs are nested on nodes. This type of structure occurs in nested state-charts, for example. We would like to construct an editor that allows a graph to be constructed in terms of nodes and edges and then to allow a particular node to be selected by double-click so that the editor switches focus and displays the graph that is {\it on the node}. Of course this process can continue to any level of nesting. If the user wishes to move from a child graph to a parent graph then the up-arrow key is used to return from the most recent double-click.

The requirement that the editor can move from parent to a descendent at any level and then return incrementally back through the levels, implies that the editor must have a state that is used to record the current level of nesting. This can easily be achieved with a projectional editor because the abstract syntax structure can contain any amount of information that is ignored by the reduction mapping. The state of the example editor is to be represented as {\tt (machine graph dump)} where {\tt graph} is the current graph to be displayed in concrete syntax and {\tt dump} is a stack of machine states that is used when descending and ascending the nested graph structure. 
\begin{wrapfigure}[13]{r}[2pt]{2.2in}
\centering
\includegraphics[width=0.3\textwidth]{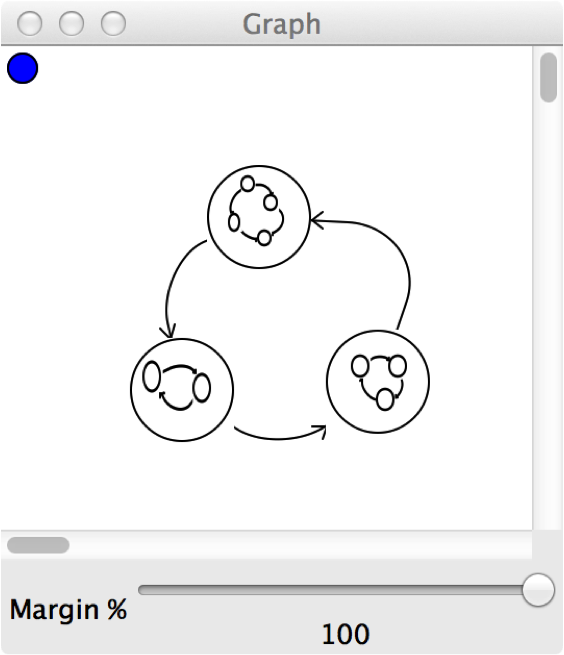}
\caption{Nested Graphs}
\label{fig:nested_graphs}
\end{wrapfigure}

A typical abstract-syntax machine-state is:
\begin{lstlisting}[numbers=none]
(machine (graph ((entity i) g1) ((entity j) g2) (relationship g1 g2)) d) 
\end{lstlisting}
for some nested graphs {\tt g1} and {\tt g2}, and some dump {\tt d}. The abstract-syntax graph is reduced to a concrete syntax-graph by mapping entities to nodes and relationships to edges. If the user double-clicks on {\tt g1} then the machine becomes:
\begin{lstlisting}[numbers=none]
(machine g1 (dump i (graph ((entity j) g2) (relationship g1 g2)) d))
\end{lstlisting} 
Note that the selected node is saved on the dump in terms of its identity and its graph. Note also, however, that the saved dumped graph omits the selected node since this must be restored later. For example, if after sveral edits the abstract-syntax tree has become:
\begin{lstlisting}[numbers=none]
(machine g1' (dump i (graph ((entity j) g2) (relationship g1 g2)) d))
\end{lstlisting} 
then pressing up-arrow results in a new machine state:
\begin{lstlisting}[numbers=none]
(machine (graph ((entity i) g1') ((entity j) g2) (relationship g1 g2)) d) 
\end{lstlisting}

\noindent 
In addition to the behaviour described above, we might want to display a graph in the editor so that thumbnails of nested graphs are shown inside the appropriate nodes. This might be used to help the user navigate through the structure. Figure \ref{fig:nested_graphs} is a screen-shot of the editor showing several nodes drawn as ellipses containing a thumbnail of the graph nested on the node.

\begin{figure}[t]
\begin{lstlisting}[deletekeywords={list}]
(deflang nested-graph
  (abstract
   [machine (machine G (empty))]
   [G (graph (entities (* entity)) (relationships))]
   [entity (entity G)])
  (transform 
   [(send (machine (graph (entities e1 ... ((entity i) g) e2 ...) r) dump) (double-click (list _ i)))
    (machine g (dump id (graph (entities e1 ... e2 ...) r) dump)) ]
   [(send (machine g (dump i (graph (entities e ...) r) d)) (key-pressed _ "up"))
    (machine (graph (entities ((entity i) g) e ...) r)  d)]
   [(send (machine (graph n (relationships r ...)) d) (new-edge type source target))
    (machine (graph n (relationships (relationship source target) r ...)) d)])
  (reduce
   [(machine g _) (->graph g)]
   [(->thumbnail (graph (entities e ... _) (relationships r ...)))
    (thumbnail 0 0 40 40 (graph (edge-types (edge (entity) (entity))) (->node e) ... (->edge r) ...))]
   [(->graph (graph (entities e ...) (relationships r ...)))
    (graph (edge-types (edge (entity) (entity))) (->node e) ... (->edge r) ...)]
   [(->node ((hole i) h)) ((node "n" i) (entity) i h)]
   [(->node ((entity i) g))
    ((node "n" i) (entity) i ((box "b" i) (align (ellipse 0 0 50 50 #f #f)) (align (->thumbnail g))))]
   [(->edge ((relationship r) source target)) ((edge r) source (none)  target (arrow))])) 
\end{lstlisting}
\caption{Nested Graphs}
\label{fig:nested_graphs}
\end{figure}

The language definition for nested graphs is shown in figure \ref{fig:nested_graphs}. The {\tt transform} clause contains rules that pocess the events that can occur on the editor:
\begin{description}
\item[{\tt (double-click i)}] This event occurs when the user selected an element. The supplied value {\tt i} is the identity of the selected element. In order to select a graph node, the node must contain a selectable element as part of its display. Boxes are selectable elements and are designated an identity {\tt (list "b" i)} where {\tt i} is the identity of the abstract-syntax entity. Therefore, the selected entity can be picked out using pattern matching as shown in the first transformation rule.
\item[{\tt (key-pressed i k)}] This event is supplied when the user presses a key and the element with identity {\tt i} is selected ({\tt -1} is supplied when nothing is selected). This is used in the second rule to pop the dump.
\item[{\tt (new-edge type source target)}] This event occurs when the user drags an edge from one node to another. If the nodes cannot be associated because there is no appropriate edge type defined then the event is not generated. Otherwise, if there are multiple edge types between the selected nodes then the user is required to choose via a menu. Once the event is generated, the {\tt source} and {\tt target} arguments are the identities of the abstract-syntax elements corresponding to the concrete-syntax nodes. The example shows a new relationship being created as a result of this event that is subsequently transformed into a concrete-syntax edge.
\end{description}
The graph example above does not contain any data on the nodes. A typical use of graphs for modelling languages is to place data on the nodes and to represent relationships between the data using edges and labels. Suppose that we want to define a class-based modelling language. A typical concrete-syntax for such a language is to show classes as boxes containing names and field information. Associations between classes are shown as edges between class-nodes.

\begin{figure}[b]
\centering
\subfigure[All Nodes Graphical]{\includegraphics[width=0.4\textwidth]{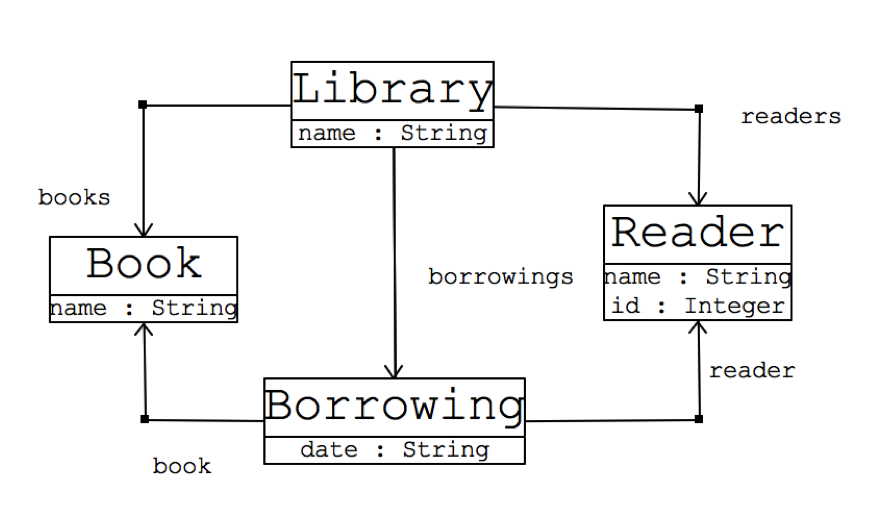}}
\subfigure[Some Nodes Textual]{\includegraphics[width=0.4\textwidth]{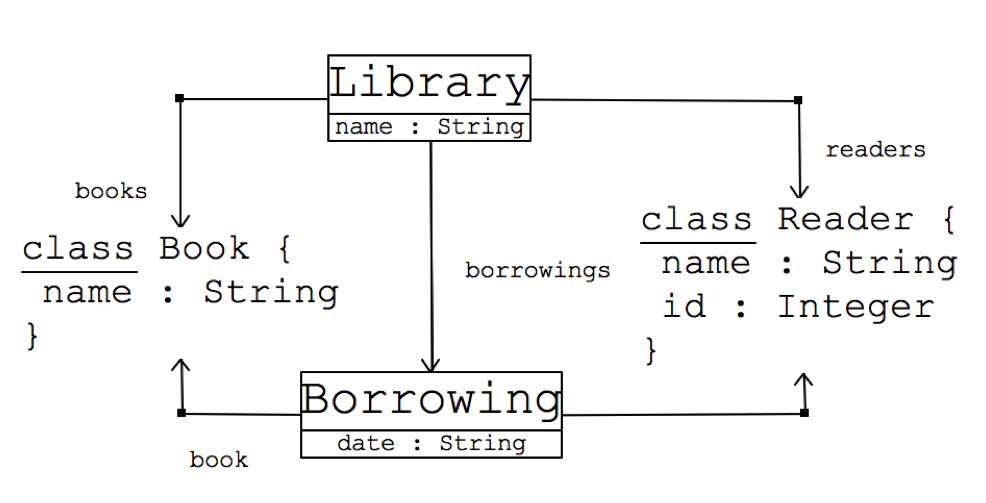}}
\caption{Mixed Mode Class Models}
\label{fig:class_models}
\end{figure}

Now suppose that we want to {\it mode} the information on the nodes by allowing the user to select the appropriate way in which the information should be displayed on a node-by-node basis. A simple example of this might be to show classes using a textual syntax. 
The resulting requirement is for a modelling language that can mix graphical with textual representations as shown in figure \ref{fig:class_models}. The screen-shot on the left shows a simple data model displayed using a conventional concrete-syntax. The screen-shot on the right shows the same model where some of the classes have been switched to a text-based representation. Although the nodes contain text, the model is still a graph that allows the nodes to be freely moved around.

\begin{figure}[t]
\begin{lstlisting}
(deflang class-models
  (abstract
   [model (model (classes (* class)) (assocs))]
   [class (class (graphical) str (* field))]
   [field (field str type)]
   [type (or (string) (integer) (boolean))])
  (transform
   [(send (model (classes c1 ... ((class i) m n f ...) c2 ...) assocs) (double-click (list _ i)))
    (change-mode ((class i) m n f ...) (classes c1 ... c2 ...) assocs)]
   [(send  (model cs (assocs a ...)) (new-edge (assoc) s t)) 
    (model cs (assocs (assoc str s t) a ...))])
  (reduce
   [(model (classes c ...) (assocs a ...))
    (graph  (edge-types (assoc (class) (class))) (class->node c) ... (assoc->edge a) ...)]   
   [(class->node ((hole i) h)) ((node "create-class" i) (new-class) i h)]   
   [(class->node ((class i) (graphical) name field ...)) (class-g-node i name field ...)]  
   [(class->node ((class i) (textual) name field ...)) (class-t-node i name field ...)] 
   [(assoc->edge ((assoc i) n sid tid)) 
    ((edge i) sid (none) tid (arrow) ((label "l" i) (target) (down-font 3 n)))]))
\end{lstlisting}
\caption{A Language for Mixed Mode Class Diagrams}
\label{fig:class_diagrams}
\end{figure}

The language definition for mixed-mode class diagrams is shown in figure \ref{fig:class_diagrams}. The language is explained as follows. The abstract-syntax definition represents a model as a collection of classes and associations. The classes can be created via a hole that will be displayed as a blue-ellipse in the editor. The associations are created via {\tt new-edge} events. Classes contain a name and a sequence of fields, and each field has a name and a type which is selected, again via a hole, from string, integer and boolean.

When a class is selected via mouse double-click, a mode change is performed using the following local:
\begin{lstlisting}[numbers=none]
[(change-mode ((class i) mode name field ...) (classes c ...) assocs)
 (case mode
   [(textual)   (model (classes ((class i) (graphical) name field ...) c ...) assocs)]
   [(graphical) (model (classes ((class i) (textual) name field ...) c ...) assocs)])]
\end{lstlisting}
When a {\tt new-edge} event occurs, a new association is created. note that the name of the association is freshly created.

Reduction of a model creates a new graph whose nodes are mapped from classes and whose edges are mapped from associations. Note that a class node might be the hole that is used to create new classes and therefore this is caught using pattern matching and translated to a node of type {\tt new-class} whose display is the hole (with the creation menu). 
A class may exist in one of two modes that affect how the nodes are created using following locals:
\begin{lstlisting}[numbers=none]
[(down-font n d) (case n [0 d] [_ (font (-) (down-font (- n 1) d))])]

[(class-g-node i name field ...)
 ((node "n" i) (class) i
  ((vbox "b" i) (fixed) (outline 1)
   (centre (up-font 3 name))
   (align ((vbox "v" i) (outline 1) (centre (down-font 3 (field-box field))) ...))))]

[(type-name t)
 (case t [(string)  "String"] [(integer) "Integer"] [(boolean) "Boolean"] [((hole _) h) h])]

[(field-box f)
 (case f
   [((field i) name type) ((hbox i) (fixed) (align name) (align " : ") (align (type-name type)))]
   [((hole i) h)          h])]

[(class-t-node i name field ...)
 ((node "n" i) (class) i
  ((box "b" i) (fixed)
   (align (tab (indent 10 
     (seq (underline "class") (space) name " {" (nl) 
          (vbox (fixed) (align (field-box field)) ...))) (nl) "}"))))]
\end{lstlisting}

\section{Implementation and Examples}

\label{sec:game_analysis}

Section \ref{sec:game} introduced a requirement for a language that could be used to both construct (model) and play (simulate) a game. The previous sections have used a variety of example languages to introduce the features of a projectional editor that are necessary to support the game. Size limitations prevent the full implementation of the game to be included in this paper, however it is included the implementation pack that accompanies the paper\footnote{\url{http://www.eis.mdx.ac.uk/staffpages/tonyclark/Software/projectional_editor_demo.zip}}. The pack includes a Mac disk image {\tt\small stand-alone-editor.dmg} of the editor implementation, several saved languages (*.xml) and the source code of the language definitions {\tt\small language-definitions.rkt}. Once you have installed the editor, navigate to the {\tt bin} directory and start the tool before dragging any of the xml files onto the editor pane to load up the language definition. The pack includes the game language definition, an example adventure, a use-cases implementation of hotel booking and a library class diagram. See the language definitions for more details.

\section{Analysis, Deficiencies and Future Directions}

This paper has introduced an architecture for projectional editors that is defined in terms of a loop processes abstract and concrete-syntax with respect to an editor that has a display and event interface. In addition the paper proposes a meta-language that can be used to initialise and drive the editor for a range of different concrete languages. The approach and meta-language has been implemented in Racket and the paper contains a number of examples that whose language definitions and screen-shots have been taken directly from the implementation.

A projectional approach allows the use of a language to be {\it modal}. This is quite easy to achieve because the mode can be a function of the abstract-syntax that can be projected onto an appropriate concrete-syntax. Examples in this paper include the class models and the game.
A declarative meta-language for abstract-syntax definition allows an editor to provide automatic support for the creation of elements and the choice between alternative elements. In the case of the game, a menu that allows new rooms to be created is auto-generated. Furthermore, by reifying the choice-point, the menus can be offered within the concrete-syntax, allowing the interactions to be context-specific.
A suitably equipped projectional editor can easily provide mixed textual and graphical interaction modes. In the case of the technology described in this paper, graphs, trees and text are all specialisations of a more general type of concrete-syntax element and can therefore be arbitrarily interleaved.

Pattern-directed mapping rules are a very convenient way of processing syntax, which is, after all, just a data structure. It is possible to identify different categories of mapping rule that are useful in this regard. Firstly, there are transformations that are used to change the state of the abstract-syntax. These are important in order to allow user-events to be handled in a consistent way. In addition, such rules can be used to add an operational semantics to a language without introducing any new features to the meta-language. 
Secondly, there are mappings that take the current abstract-syntax structure and project it onto a concrete-syntax structure that corresponds to the display interface of an editor. as has been shown, these can be achieved using pattern-directed rules and therefore require no new meta-language machinery. The range of concrete-syntax elements will depend upon the application and the sophistication of the editor, however the selection shown in this paper seems sufficient for a wide range of language-applications and could easily be extended.

Identity-management is an important feature of projectional-editing; it is required to manage the association between abstract and concrete-syntax so that events occurring in the context of a concrete-element can be associated with a corresponding abstract-element. Secondly, there is information relating to the concrete-syntax that has no counterpart in the abstract-syntax, for example the position of graph-nodes or the number of way-points on an edge. In principle, the concrete-syntax is re-generated from the abstract-syntax each time the user performs an edit and therefore, if the identities are not managed appropriately the layout information would be lost.

Although the approach and corresponding implementation have been shown to support a range of interesting and useful languages, a number of deficiencies remain to be addressed. A criticism of projectional technology is that it tends to be syntax-directed which can reduce usability by forcing the user to enter language structures through a pre-defined interface. One possibility is to take a light-touch approach whereby the user is only aware of the structure behind the concrete-syntax when they either ask to be told or when it becomes helpful to do so. To the user it appears as though they are working with a text-based editor with a knowledgeable expert looking over their shoulder occasionally prompting them or to whom they can refer when things get tricky. 
This approach is fine when dealing with certain types of language, for example programming languages. However, projectional technology is arguably not the best support for programmers whose mental models of their systems are (or ought to be) far more sophisticated, dynamic and responsive than any projectional technology could ever hope to be. Perhaps projectional technologies might best be applied in cases where there is obviously a need for structural support, for example when zooming in and out of structures that need to be related or when there is a repeated pattern of structure to be completed. In addition, it is possible that different styles of editor-interaction are required within the same system development: projectional-style for the big-picture stuff and more traditional-style (albeit with knowledgeable support) for the detail. In any case, it is not clear which approach is more appropriate and the case has not been made that any one approach is universal, therefore this paper is claimed to be a valid contribution to the debate. 

One possible future area of investigation is to use grammar-ware at the leaves of the syntax-tree and to integrate both free-text editing and syntax-directed editing at that level. A fruitful strategy might be to allow text-boxes to be associated with grammars that synthesize abstract-syntax. If the grammar can be successfully used to parse the text then the result is subject to projectional mapping. Otherwise the content remains free text. This strategy would support mode-shifts that address the oft cited problem of modifying the internal leaves of a binary expression tree by inserting new operators without resorting to syntax-directed menus.

The meta-language described in this paper provides no support for error handling and will simply go wrong if the rules fail to produce a normal-form or if a local rule definition produces a tree of an unexpected type. One way to address this is to have a separate category of rules that are used for checking and error reporting. If the editor provides a separate error interface that can be used for bringing error messages to the attention of the user and for linking the errors to the concrete syntax then the same pattern-based rule technology can be used to perform error handling.
Related to this, the language does not support static checking. For example, it should be possible to detect the use of unbound identifiers and undefined functors. Some aspects of static checking should be easy to achieve, however it would also be desirable to define a type system so that the use of syntax structures can be checked before use. It is not clear whether it is possible to determine if the reduction rules can be shown to terminate with a valid normal-form, or under what conditions this is possible. 

There is interest in the modularity and composition of languages and DSLs in particular  \cite{tomassetti2013model,white2009improving,cazzola2010dsl,cazzola2009sectional,renggli2010language,rumpe2013towards}. A key challenge to achieving engineered integration is posed by concrete-syntax. By inverting the focus of attention to abstract-syntax, a projectional editor does not suffer from such problems. However, there are still significant issues to be addressed and this could be a fruitful area for future work.

The current implementation has an XML-based save and load mechanism that is based on an obvious encoding of the syntax trees defined in figure \ref{fig:editor_architecture}. Given the universality of the data representation it should be possible to import data in any well-defined format, such as XMI or EMF, in order to create tool-chains.

\bibliographystyle{plain}
\bibliography{refs}

\end{document}